\documentstyle[preprint,aps]{revtex}
\begin{document}
\draft
\preprint{}
\title
{A HEAVY HIGGS CAN GIVE STRONG FIRST ORDER ELECTROWEAK PHASE TRANSITION
IN THE STANDARD MODEL $?$}
\author
{B. B. Deo and L. Maharana }
\address
{Physics Department, Utkal University, Bhubaneswar-751004, India.}
\maketitle
\begin{abstract}
The role of Higgs sector of the standard model in first order 
phase transition is reexamined.It is found that a solution to the
mass gap equations exist which can be used in higher orders.
This possible solution leads to a transition which is found to be
strongly first order and the ratio of the critical scaler field
and the critical temperature is about 1.2 for a Higgs of mass
around 400 GeV.This can explain baryogenesis.

\end{abstract}

Recently electroweak phase transition at high temperature
has been a topic of great interest. Sakharov \cite{ref1}
was the first to point out that the observed baryon asymmetry
of the Universe can be produced by processes which violate
C, CP and B and occurs out of thermal equilibrium. All the conditions
can be met by the Standard Model; C-violation exists, CP violating terms 
can be accomodated, Sphaleron can induce sufficient B-violating processes 
at high temperature. A first order phase transition \cite{ref2} can provide
the nonthermal equilibrium. In order to have sufficient departure from 
equilibrium, it is necessary that this transition be rather strong.
The value of the ratio of the Higgs field ${\Phi}_c$ and the critical 
temperature $T_c$ should at least be about 1.0 . All effective potentials
at high temperatures constructed so far obtain this ratio to be
rather small. Furthermore it has been shown to decrease when Higgs mass 
is increased \cite{ref3,ref4,ref5}. In this letter we propose to construct
a plausible theory by solving the gap equation of
Ref.[6] to the required order.The ratio 
${\Phi}_c/T_c$ increases with the value of the mass of the Higgs
and has the value of 1.0 when the Higgs mass is round 400 GeV.

The most basic quantity to calculate is the one loop effective potential
$V_{eff}(\Phi,T)$, improved by appropriate 
resummation \cite{ref5a}  and two loop calculations. We assume that 
this potential can be cast into the form
\begin{equation}
V_{eff}(\Phi,T)=a {\Phi}^2-b{\Phi}^3+c{\Phi}^4\;,
\end{equation}
plus a constant field independent term which shall be omitted.
a, b and c are functions of the temperature. $V_{eff}(\Phi,T)$ should be
real and imaginary parts possibly do not arise for high temperature.
The effective Higgs and Goldstone field dependant masses are
\begin{equation}
{m_{eff}}^{2}_{Higgs}(\Phi,T)=\frac{{\partial}^2V(\Phi,T)}{{
\partial \Phi}^2}\;
\end{equation}
and
\begin{equation}
{m_{eff}}^{2}_{Goldstone}(\Phi,T)=\frac{1}{\Phi} \frac{\partial 
V(\Phi,T)}{\partial \Phi}\;.
\end{equation}
The potential should be extremum at the origin.

The classical potential is

\begin{equation}
V_0=\frac{\lambda}{4}({\Phi}^2-{\sigma}^2)^2\;,
\end{equation}
where $\lambda$ is the coupling constant and $\sigma$=246 GeV
is the symmetry breaking value of the field. The effective field 
dependent masses of particles in the standard model are, in obvious
notations,
\begin{equation}
{m_H}^2(\Phi)=3 \lambda {\Phi}^2-\lambda {\sigma}^2\label{eq:1},
\end{equation}

\begin{equation}
{m_G}^2(\Phi)= \lambda {\Phi}^2-\lambda {\sigma}^2\label{eq:2},
\end{equation}

\begin{equation}
{m_w}^2(\Phi)=\frac{{M_w}^2}{{\sigma}^2} {\Phi}^2,
\end{equation}

\begin{equation}
{m_t}^2(\Phi)= \frac{{M_t}^2}{{\sigma}^2} {\Phi}^2
\end{equation}
and
\begin{equation}
{m_z}^2(\Phi)= \frac{{M_z}^2}{{\sigma}^2} {\Phi}^2\;.
\end{equation}
We shall omit, for this theory, the contribution of fermions
lighter than the top. The zero temperature effective potential
as modified by radiation correction up to one loop is \cite{ref3,ref4}.

\begin{eqnarray}
V^{(1)}(\Phi,0)=&&\sum_{i=H,w,z,t}\frac{n_i}{64{\pi}^2}\left [
{m_i}^4(\Phi)(log\frac{{{m_i}^2(\Phi)}}{{M_i}^2}-\frac{3}{2})
+2{m_i}^2(\Phi){M_{i}}^2 \right ]\nonumber\\
&&+\frac{n_G}{64{\pi}^2}{m_G}^4(\Phi)(log \frac{{m_G}^2(\Phi)}{{M_H}^2}
-\frac{3}{2})\;,
\end{eqnarray}

where $n_H$=1, $n_G$=3, $n_w$=6, $n_z$=3, $n_t=-$12.
At finite temperature, the one loop effective potential is

\begin{equation}
V_T=\frac{T^4}{2{\pi}^2}\left [ \sum_{i=H,G,W,Z} n_i J_+({y_i}^2)
+ n_t J_-({y_t}^2)\right ],
\end{equation}
where ${y_i}^2=\frac{{m_i}^2(\Phi)}{T^2}$,
\begin{equation}
J_{\pm}(y_i^2)={{\int}_0}^{\infty}dx x^2 log[1{\mp}exp(-\sqrt{x^2+y^2})]\;.
\end{equation}

It has been shown by Anderson and Hall \cite{ref5} that for values of
$\frac{{m_i}(\Phi)}{T}$ less than 1.6, the high temperature approximation,
namely

\begin{eqnarray}
V_T(\Phi,T) \simeq && \sum_{i=H,G,W,Z} n_i 
\left [ \frac{{{m_i}^2(\Phi)} T^2}{24}
-\frac{{m_i}^3(\Phi)}{12 \pi}T
-\frac{{{m_i}^4(\Phi)}}{64{\pi}^2}(log\frac{{m_i}^2(\Phi)}
{T^2}-5.4076)\right ]\nonumber\\
&&-n_t \left [ \frac{{{m_t}^2(\Phi)} T^2}{48}+\frac{{m_t}^4(\Phi)}{64{\pi}^2}
(log\frac{{m_t}^2(\Phi)}{T^2}-2.6350)\right ], 
\end{eqnarray}
is valid to better than $5\%.$ Our values for the ratio will be less than 1.6
even if the mass of the Higgs will be very high. The $m^4$ log${m^2}$ 
term of the T=0 term cancells with the term T$\neq$0, leaving the quartic 
mass term with a constant temperature dependent factor.

There are important corrections due to lollipops, daisies and superdaisies
\cite{ref6}. We shall consider only few of them. The 
contribution to the ${\Phi}^3$ term in the gauge sector is reduced by
a factor of $\frac{2}{3}$.
Since gauge couplings are small, we shall completely neglect the
higher order corrections due to them. Reference [3] treats
the gauge sector very thoroughly. The important contribution
that really matters is a log($\Phi$) term \cite{ref9} which  does
not alter our results.\\ 
There are actually three types of expansions, the perturbative expansion in
$\lambda$ at zero temperature, the high temperature expansion in
$\frac{m^2(\phi)}{T^2}$ and the loop expansion which we shall discuss later.
At zero temperature the perturbation theory is valid upto $\lambda $=3.5
or thereabout~\cite{ref4a}. Due to lollipops, daisies and superdaisies the 
contributions to the effective potential from the Higgs sector only, has
been calculated by many authors. From the results of Boyd, Brahm and Hsue
\cite{ref4}, it is easily seen that at high temperatures the quadratic
part of the potential is enhanced by the addition of $\lambda \phi^2 T^2$ 
due to Higgs. Consistent with high temperature expansion limiting values
and to the first order in $\lambda$ we now collect $\phi^4$,$\phi^2 T^2$
terms and cubic gauge contribution term $\phi^3 T$. The cubic term of the
Higgs sector is left for the detail consideration later. 
The result is

\begin{equation}
V(\Phi,T)=\frac{1}{4} {\lambda}_T {\Phi}^4+d (T^2-{T_0}^2){\Phi}^2
-e{\Phi}^3 T+V^{(3)}(\phi,T)
\end{equation}

\begin{eqnarray}
{\lambda}_T=&& -\frac{3}{16{\pi}^2{\sigma}4} [ {2M_w}^4(log\frac{{M_w}^2}
{T^2}-3.91)+{M_z}^4(log\frac{{M_z}^2}{T^2}-3.91)\nonumber\\
-&&4{M_t}^4(log\frac{{M_t}^2}{T^2}-1.14)+{M_H}^4(log\frac{{M_H}^2}{T^2}
-3.91) ],
\end{eqnarray}
\begin{equation}
d=\frac{1}{8{\sigma}^2}({M_H}^2+2{M_w}^2+{M_z}^2+2{M_t}^2)\;,
\end{equation}
\begin{equation}
{{T_0}^2}=\frac{6{M_H}^2-\frac{3}{4{\pi}^2{\sigma}^2}
(1.5{M_H}^4+6{M_w}^4+3{M_z}^4-12{M_t}^4)}{3{M_H}^2+6{M_w}^2+3{M_z}^2+6{M_t}^2}
{\sigma}^2,
\end{equation}
\begin{equation}
e={e_g}+{e_H},
\end{equation}
\begin{equation}
{e_g}=\frac{1}{6 \pi {\sigma}^3} (2{M_w}^3+{M_z}^3)\simeq {10}^{-2}
\end{equation}
and
\begin{equation}
V^{(3)}=-{T\over {12\pi}}(m_H^3(\phi)+3m_G^3(\phi)).
\end{equation}
$e_H$ is the contribution from the Higgs sector. We replace $\lambda$ by
$\lambda_T$ everywhere which we assume to be a partial resummation.
The most difficult part has been to deal with the cubic terms of the
Higgs and Goldstone sectors. In the lowest order both $m_H$ and $m_G$
become imaginary. 
Most of the earlier works have mainly taken the approximate values
$m_H^2=3 \lambda \phi^2+2d(T^2-T^2_0)$ and $m_G^2=\lambda \phi^2+
2d(T^2-T^2_0)$. We want to find the value of $V^{(3)}$ by following an 
entirely different method.
We first calculate the effective masses
\begin{equation}
{m_{H,eff}^2}(\Phi,T)=3{\lambda}_T {\Phi}^2 + 2 d(T^2-{T_0}^2)
-6e\Phi T+\frac{{\partial}^2 V^{(3)}}{\partial\phi^2}\label{eq:e1}
\end{equation}
and
\begin{equation}
{m_{G,eff}^2}(\Phi,T)={\lambda}_T{\Phi}^2 + 2 d(T^2-{T_0}^2)
-3e\Phi T+\frac{1}{\phi} \frac{\partial V^{(3)}}{\partial\phi}.\label{eq:e2}
\end{equation}
Let us compare this with the gap equation of Buchmuller etal \cite{ref7}
.To the order ${\lambda}^{3/2}$ and $g^2$, they are in our notation,
\begin{eqnarray}
{m_{H}^2}= && 2 d(T^2-{T_0}^2)+3{\lambda}_T{\Phi}^2 -6e_g\Phi T\nonumber\\
&&-\frac{3{\lambda}_T}{4\pi} [ m_H+m_G+{\lambda}_T{\Phi}^2(\frac{3}{m_H}
+\frac{1}{m_G}) ] T\label{eq:e2a}
\end{eqnarray}
and
\begin{eqnarray}
{m_{G}^2}= && 2 d(T^2-{T_0}^2)+{\lambda}_T{\Phi}^2 -3e_g\Phi T\nonumber\\
&&-\frac{{\lambda}_T}{4\pi}\left [m_H+5m_G+4{\lambda}_T{\Phi}^2(\frac{1}{{m_H}+
{m_G}})\right ]T.\label{eq:e2b}
\end{eqnarray}
The last terms of these expressions are $\lambda^{3\over 2}$ corrections 
and can be related to
the cubic mass terms of the full potential. Using the equations (\ref{eq:1})
and (\ref{eq:2})
and comparing with the gap equations (\ref{eq:e2a}) and (\ref{eq:e2b}) we
obtain

\begin{eqnarray}
\frac{\partial^2 V^{(3)}}{\partial \phi^2}=&&
-{T\over {12\pi}} \{ [6m_H(\phi)(\frac{\partial m_H(\phi)}{\partial {\phi}}
)^2+3m_H^2(\phi)(\frac{\partial^2m_H(\phi)}
{\partial \phi^2})] \nonumber\\
&&+3[6m_G(\phi)(\frac {\partial m_G(\phi)}{\partial \phi})^2 +3m_G^2(\phi)
\frac{\partial^2m_G(\phi)}{\partial \phi^2}] \}\nonumber\\
&&=-{{3{\lambda}_T}\over {4\pi}}T
(m_H(\phi)+m_G(\phi)+{\lambda}_T{\phi}^2({3\over m_H(\phi)} 
+{1\over m_G(\phi)}))\label{eq:ee2}
\end{eqnarray}
and
\begin{eqnarray}
\frac{1}{\phi} \frac{\partial V^{(3)}}{\partial\phi}=&&
-{T\over {12\pi\phi}} [3m_H^2(\phi)
(\frac {\partial m_H(\phi)}{\partial \phi})
+9m_G^2(\phi)\frac {\partial m_G(\phi)}{\partial \phi}]\nonumber\\  
&&=-{{{\lambda}_T}\over {4\pi}}T
(m_H(\phi)+5m_G(\phi)+4{\lambda}_T{\phi}^2
({1\over {m_H(\phi) + m_G(\phi)}}))\label{eq:ee3}
\end{eqnarray}
Our main interest is to examine the possibility of a strong first order
phase transition. Interestingly we find that these equations have a unique
property. If $m_G$ and $m_H$ are proportional to $\sqrt{\lambda} \phi$, the 
dependance of the equations on $\lambda$ and $\phi$ cancel out. The cubic term
is of order $\lambda^{3\over 2}$ like the correction terms of the gap
equation. Let us take $m_H$=a$\sqrt{\lambda}\phi$ and  
$m_G$=b$\sqrt{\lambda}\phi$. The equations (\ref{eq:ee2}) and (\ref{eq:ee3})
reduces to
\begin{equation}
2a^3+6b^3+3\lambda (a+b)+3{\lambda}^2 (\frac{3}{a}+\frac{1}{b})=0
\end{equation}
and
\begin{equation}
a^3+3b^3-\lambda (a+5b)-\frac{4\lambda^2}{a+b}=0.
\end{equation}
Using Newton-Ralphson method we find there are several solutions. Ones that
interest us is a=1.732=$\sqrt{3}$ and b=1. So in evaluating the cubic 
term of the potential $V^{(3)}$ we shall take $m_H=\sqrt{3\lambda}\phi$ and
$m_G=\sqrt{\lambda}\phi$. The value of $e_H$ comes out to be
\begin{equation}
e_H=\frac{\sqrt{3}+1}{4\pi} \lambda^{3\over 2}_T
\end{equation}
and
\begin{equation}
V^{(3)}=-(e_g+e_H)\phi^3 T=-e\phi^3 T.
\end{equation}
The idea of a strong first order phase transition has been previously
ignored. The reason that has been advanced is that the perturbation
theory fails at high temperatures for large values of $\lambda$.
Let us examine the loop expansion parameter following Buchmuller
et al. Inspection of gap equation (\ref{eq:e2a}) and (\ref{eq:e2b}) 
shows that if we introduce expansion parameter by 
\begin{equation}
\zeta_H \frac{\lambda_T T}{4 \pi}\left (\frac{3}{m_H}+\frac{1}{m_G}
\right )=1
\end{equation}
\begin{equation}
\zeta_G \frac{\lambda_T T}{\pi}\left (\frac{1}{m_H+m_G}
\right )=1\;,
\end{equation}
the largeness of $\zeta_H$ and $\zeta_G$ will imply convergence
of the perturbative expansion. We have introduced the parameter
$\zeta_G$, coming from equation (23) as the effective potential can be
determined by integrating $\Phi m_G^2(\Phi)$.

The occurance of $m_H$ and $m_G$ imply that the expansion will fail down
when they will be imaginary. From equations (24) and (20) it is
seen that $m_H^2$ will be negative when $3{\lambda}{\Phi}^2+2d(T^2-T_0^2)$
becomes less than $6e \Phi T$. This happens at a value $\Phi=\Phi_b$
and $T=T_b$ which can be obtained by completing square as follows :

\begin{equation}
m_H^2=3\lambda (\Phi-\frac{e}{\lambda}T)^2-3\frac{e^2}{\lambda}T
+2d(T^2-T_0^2)\;.
\end{equation}
This give
\begin{equation}
\frac{T_b^2-T_0^2}{T_b^2}=\frac{3}{2}\frac{e^2}{\lambda d}\;\;\;\;\;
\mbox{and}\;\;\;\;\;\Phi_b=\frac{e}{\lambda}T_b\;.
\end{equation}
At this temperature $T_b$, following Buchmuller et al.
\begin{equation}
\bar{m}_H \simeq \sqrt{6 \frac{e^2}{\lambda_T}} T_b
\end{equation}
and    
\begin{equation}
\bar{m}_G \simeq \sqrt{3 \frac{e^2}{\lambda_T}} T_b
\end{equation}
Since $e\simeq e_H$, the expansion parameters are obtained as
\begin{equation}
\zeta_H=(\sqrt{3}+1)/(\sqrt{\frac{3}{2}}+\frac{1}{\sqrt{3}})=1.52
\end{equation}
\begin{equation}
\zeta_G=(\sqrt{3}+1)(\sqrt{3}+\sqrt{6})/4=2.856
\end{equation}
Both the expansion parameter come out independent of the Higgs mass.
The convergence of the perturbative expansion is slow but sure and
better if effective potential is calculated by integrating
$\Phi m_G^2(\Phi)$.
We can now write the effective potential as
\begin{equation}
V(\Phi,T)=\frac{{\lambda}_T}{4}{{\Phi}^4}+d (T^2-{T_0}^2)\phi^2-e{\Phi}^3T.
\end{equation}
Here
\begin{equation}
e=e_g+\frac{3\sqrt{3}+3}{12\pi}{{\lambda}_T}^{3/2}.
\end{equation}
We calculate $V(\Phi,T)$ and find the critical temperature which is given
in Table I. The graphs at the critical temperature $T_c$ are plotted in 
Fig.1(a) and (b).

It is to be noted that value of 
$({\lambda}_T-\lambda )/{\lambda}_T$ is found 0.3 for all cases 
considered. We must now examine the validity of the high 
temperature expansion. 
In general the equation for $V(\phi,T)$can be recast in the form
\begin{equation}
V(\Phi,T)=\frac{{\lambda}_T}{4}{{\Phi}^4}+d (T^2-{T_0}^2){\Phi}^2
-eT{\Phi}^3.
\end{equation}
In the above quantities like d, $T_0$ and e become mildly temperature 
dependent. At the critical temperature the following equations hold:
\begin{equation}
{T_c}^2=\frac{{T_0}^2}{1-(e^2/{\lambda}_Td)}\;\;\;
and\;\;\;\;\frac{{\Phi}_c}{T_c}=2\frac{e}{{\lambda}_T}.
\end{equation}

The highest mass term is to the effective Higgs mass which from the full
potential is
\begin{equation}
{m_{H,eff}}^2(\Phi,T)=3 {\lambda}_T{\Phi}^2+2dT^2-{T_0}^2)-6e\Phi T 
\end{equation}
and at $T=T_c$
\begin{equation}
{m_{H}^2}({\Phi}_c,T_c)=\frac{ {\lambda}_T}{2}{{\Phi}_c}^2 
\end{equation}
and
\begin{equation}
\frac{m(\phi)_{\phi=\phi_c}}{T_c}=
\sqrt{\frac{{\lambda}_T}{2}} \frac{{\Phi}_c}{T_c} \simeq 1.17
\end{equation}
at $M_H$=425 GeV.
For $M_H=425$ GeV \cite{ref8}  $\lambda =1.49$ and ${\lambda}_T=2.13$.
This is about the limit where the high temperature expansion holds 
to better than 5\% \cite{ref5}. 

An increase in Higgs mass raises the important ratio 
${\Phi}_c/T_c$. The Higgs mass, of course, can not be arbitrarily large as
the high temperature expansion will breakdown possibly
for Higgs mass above 500 GeV \cite{ref10}.

Thus we have shown that contrary to current results in literature,
the first order phase transition can be strong and gets stronger with increase
of Higgs mass. It is thus possible to explain Baryogenesis with a very 
massive Higgs. 

The perturbation theory at zero temperature seems to be valid for 
Higgs mass below 700 GeV \cite{ref10}. However at high temperature with the gap
solution conventionally taken , as has been done by Buchmuller et al, the upper
limit of Higgs mass is found to be 70 GeV . From the 
solution of gap equation chosen here the loop expansion parameter at
high temperature is independent of Higgs mass.

The result of lattice simulations \cite{ref11} are not conclusive. Some of them
e.g. Bunk et al and Kajantie et al have concluded that for intermediate
Higgs mass, there is strongly first order phase transition. Moreover all 
lattice calculations have ignored the fermions which must be playing a role
in cancellation of vacuum loops of Higgs sector. A comprehensive analysis for
Higgs mass including top is still wanting. In any event, the lattice 
calculations have not reached any definite conclusions.

The purpose of this letter is to emphasize the role of the Higgs 
sector which not only breaks the $SU(2)$ symmetry but may give a strong
first order electroweak phase transition.

One of the authors (B.B.D.) is thankful to the University 
Grants Commission for supporting a research project.

\begin{table}
\caption{Values of $\lambda$,${\lambda_T}$,$\phi_c$,$T_c$
and $\frac{{\phi}_c}{T_c}$ against $m_H$}
\begin{tabular}{cccccc}
\multicolumn{1}{c}{$m_H$(GeV)}&\multicolumn{1}{c}{$\lambda$}
&\multicolumn{1}{c}{${\lambda}_T$}&\multicolumn{1}{c}{$\phi_c$
(GeV)}&\multicolumn{1}{c}{$T_c$(GeV)}&\multicolumn{1}{c}
{$\frac{\phi_c}{T_c}$}\\
\tableline
 150&0.185&0.179&77&164&0.46\\
 200&.33&.337&88&201&.438\\
 250&.51&.56&110&232&.47\\
 300&.74&.86&152&263&.58\\
 350&1.01&1.25&215&298.9&.72\\
 400&1.32&1.76&339&362&.93\\
 425&1.49&2.13&540&447&1.13\\
\end{tabular}
\end{table}
\end{document}